\documentclass[]{spie}  

\usepackage{amsmath,amsfonts,amssymb}
\usepackage{graphicx}
\usepackage[colorlinks=true, allcolors=blue]{hyperref}

\title{QSNET, a network of clocks for measuring the stability of fundamental constants}

\author[a]{G. Barontini}
\author[a]{V. Boyer}
\author[b]{X. Calmet}
\author[c]{N. J. Fitch}
\author[a]{E. M. Forgan}
\author[d]{R. M. Godun}
\author[a]{J. Goldwin}
\author[a]{V. Guarrera}
\author[d]{I. R. Hill}
\author[a]{M. Jeong}
\author[b]{M. Keller}
\author[b]{F. Kuipers}
\author[d]{H. S. Margolis}
\author[a]{P. Newman}
\author[a]{L. Prokhorov}
\author[c]{J. Rodewald}
\author[c]{B. E. Sauer}
\author[d]{M. Schioppo}
\author[b]{N. Sherrill}
\author[c]{M. R. Tarbutt}
\author[a]{A. Vecchio}
\author[a]{S. Worm}

\affil[a]{School of Physics and Astronomy, University of Birmingham, Edgbaston, Birmingham B15 2TT, UK}
\affil[b]{Department of Physics and Astronomy, University of Sussex, Brighton BN1 9QH, UK}
\affil[c]{Centre for Cold Matter, Blackett Laboratory, Imperial College London, Prince
Consort Road, London SW7 2AZ UK}
\affil[d]{National Physical Laboratory, Hampton Road, Teddington TW11 0LW, UK}

\authorinfo{Further author information: (Send correspondence to G. B.)\\G. B.: E-mail: g.barontini@bham.ac.uk}

\begin{document} 
\maketitle

\begin{abstract}
The QSNET consortium is building a UK network of next-generation atomic and molecular clocks that will achieve unprecedented sensitivity in testing variations of the fine structure constant, $\alpha$, and the electron-to-proton mass ratio, $\mu$. This in turn will provide more stringent constraints on a wide range of fundamental and phenomenological theories beyond the Standard Model and on dark matter models.  
\end{abstract}

\keywords{variations of fundamental constants, atomic and molecular clocks, networks of quantum sensors, dark matter, dark energy, physics beyond the Standard Model}

\section{Introduction}
\label{sec:intro}
QSNET is aiming to create a world-leading programme to search for spatial and temporal variations of fundamental constants using high-precision spectroscopy in networked quantum clocks. The network will include existing Sr, Yb$^+$ and Cs atomic clocks at the National Physical Laboratory in London and  several new clocks that are currently being developed: a N$_2^+$ molecular ion clock at the University of Sussex, a CaF molecular optical lattice clock at Imperial College London, and a Cf highly-charged ion clock at the University of Birmingham. As the programme progresses, QSNET can be expanded and linked with other clocks across the globe. An important objective of QSNET is to exploit the networked approach, in which multiple quantum sensors will be linked, having complementary sensitivities to changes of fundamental constants. In particular, QSNET will achieve unprecedented accuracy in testing variations of the fine structure constant, $\alpha$; and the electron-to-proton mass ratio, $\mu$. Such variations could manifest as slow drifts, oscillations or transient events, and all of them may be detectable by QSNET. This will allow us either to discover new physics or to provide tighter constraints on e.g. specific dark matter and dark energy models, soliton models and violations of fundamental symmetries. 

\section{Theoretical background}
\label{sec:theory}  

Fundamental constants of nature are by definition assumed to be immutable in time or space. However, since the late 1930's, there have been speculations  \cite{dirac1,dirac2,Milne,Jordan,Landau}, starting with Dirac's large numbers hypothesis, that fundamental constants could vary in time. The interest in such a time evolution was revived some 20 years ago by astrophysical observations \cite{Webb:2000mn} claiming a potential discovery that the fine structure constant was smaller in the past at high redshift. However, this observation is not supported by other groups see e.g.  Chand et al. \cite{Chand:2004ct} and the situation remains open. A key issue with astrophysical observation is that it is difficult to control all the parameters. This is a very strong motivation to look for such effects using earth based experiments. While astrophysical observations can probe effects of time variation which accumulate over long periods of time i.e. some billions of years (corresponding to redshifts of 3 to 5), measurements with clocks can only be performed over periods of time corresponding to days or years. However, clocks clearly surpass any astrophysical measurement of a potential time evolution of fundamental constants in terms of precision. The unrivalled precision of clocks provides us with an exceptional tool for measuring variations of fundamental constants.

There are many theoretical motivations to consider a possible time evolution of fundamental constants, see e.g.  the review by Uzan  \cite{Uzan:2010pm}. Models can range from those with extra-dimensions, to quintessence models and models of extremely light dark matter.  A time dependence of the coupling constants of the Standard Model of particle physics can be parametrized by a scalar field $\phi$ which couples to the electron $\psi_e$, light quarks ($u$, $d$ and $s$-quarks) $\psi_q$, the photon $A_\mu$ or gluons $G^a_\mu$ according to
\begin{eqnarray}
	L&=&\kappa  \phi \left (\frac{d^{(1)}_e}{4} F_{\mu\nu}F^{\mu\nu} -d^{(1)}_{m_e} m_e \bar \psi_e \psi_e \right ) 
	+ \kappa  \phi \left (\frac{d^{(1)}_g}{4} G_{\mu\nu}G^{\mu\nu} -d^{(1)}_{m_q} m_q \bar \psi_q \psi_q \right )
\end{eqnarray}
with $\kappa=\sqrt{4 \pi G_N}$,  $F_{\mu\nu}=\partial_\mu A_\nu-\partial_\nu A_\mu$ and $G_{\mu\nu}=\partial_\mu G_\nu-\partial_\nu G_\mu-i g_s [G_\mu,G_\nu]$ where $g_s$ is the QCD coupling constant and $G_N$ is the Newtonian constant of gravitation. We could also add couplings to the field strength of the neutrinos, heavier leptons and quarks, electroweak gauge bosons of the standard model and to the Higgs bosons, but these particles usually do not play an important role for very low energy tabletop experiments such as clocks, so can be ignored at this stage.  Note that these operators are dimension 5 operators as they are suppressed by one power of the reduced Planck scale $M_P=1/\sqrt{8 \pi G_N}$. 

For some applications, it might be necessary to consider scalar fields that transform under some discrete, global or gauge symmetry in which case, the simplest coupling to matter is given by dimension 6 operators 
\begin{equation}
	L=\kappa^2  \phi^2 \left (\frac{d^{(2)}_e}{4} F_{\mu\nu}F^{\mu\nu} -d^{(2)}_{m_e} m_e \bar \psi_e \psi_e \right ) 
	+ \kappa^2  \phi^2 \left (\frac{d^{(2)}_g}{4} G_{\mu\nu}G^{\mu\nu} -d^{(2)}_{m_q} m_q \bar \psi_q \psi_q \right ).
\end{equation}
In other words, the interactions of the scalar field with stable matter are suppressed by two powers of the reduced Planck scale. These are non-linear couplings. 

These simple Lagrangians can account for a variety of physical phenomena which can be probed with the QSNET network, e.g. scalar dark matter models in which case the magnitude of $\phi$ is related to the density of dark matter, quintessence-like models \cite{Dvali:2001dd}, generic hidden sector scalar field \cite{Calmet:2019frv}, Kaluza-Klein models \cite{Marciano:1983wy}, dilaton field models or Brans-Dicke fields, transient phenomena, cosmic strings, and domain walls. 

One expects on very general grounds that quantum gravity will generate an interaction between any scalar field $\phi$ and regular matter with $d_j^{(i)}\sim {\cal O}(1)$ whether such a coupling exists or not when gravity decouples \cite{Calmet:2019jyz,Calmet:2019frv,Calmet:2020rpx,Calmet:2020pub,Calmet:2021iid}. However, very light scalar fields coupling linearly to regular matter (i.e. dimension five operators) are essentially ruled out by the E\"ot-Wash torsion pendulum experiment \cite{Kapner:2006si,Hoyle:2004cw,Adelberger:2006dh,Lee:2020zjt}  for $d_j^{(1)}\sim {\cal O}(1)$. Indeed, E\"ot-Wash's data implies that if  $d_j^{(1)}\sim1$, the mass of the singlet scalar field must be heavier than $10^{-2}$ eV, and thus to be relevant for clocks  $d_j^{(1)}\ll1$\cite{Calmet:2019jyz,Calmet:2019frv,Calmet:2020rpx,Calmet:2020pub,Calmet:2021iid}.  QSNET will therefore provide a very important test of quantum gravity \cite{Calmet:2019jyz,Calmet:2019frv,Calmet:2020rpx,Calmet:2020pub,Calmet:2021iid}. If a very light neutral scalar field with linear coupling to regular matter was found with QSNET, we would learn that dimension 5 operators are not generated by quantum gravity.  On the other hand, non-linear couplings are far less constrained by current experiments and clocks will be able to explore uncharted territory. 

Besides quantum gravity, QSNET enables tests of grand unified theories \cite{Calmet:2001nu,Calmet:2002ja,Calmet:2002jz,Langacker:2001td,Campbell:1994bf,Olive:2002tz,Dent:2001ga,Dent:2003dk,Landau:2000cc,Wetterich:2003jt,Flambaum:2006ip,Calmet:2014qxa}, as it allows us to measure at the same time variations of $\alpha$ and $\mu$ using different clocks of the network. In grand unified models, shifts in $\alpha$ and $\mu$ are related and the functional dependence is very model dependent. QSNET could therefore discriminate between models that predict a variation of the unified coupling constant or a time-variation of the unification scale, or both. Finally, QSNET can probe space-time symmetries such as Lorentz invariance, CPT, and also probe models of space-time non-commutativity \cite{Calmet:2004dn,Carlson:2001sw}.  QSNET thus enables tests of cosmology, astrophysics and particle physics in a laboratory with tabletop experiments.

\section{The QSNET network}
\label{sec:exp}  

QSNET will include a range of clocks selected to maximize the sensitivities $K_{\alpha}$ and $K_{\mu}$ to the variation of $\alpha$ and $\mu$. Here, we summarize the clocks and give their sensitivity factors in atomic units.

\begin{itemize}
\item A $^{133}$Cs microwave fountain clock. This clock is sensitive to changes in both the fine structure constant ($K_{\alpha} = 2.83$) and the electron-to-proton mass ratio ($K_{\mu} = 1$). There are several state-of-the-art Cs fountain clocks around the world~\cite{Weyers2018,Heavner2014,Guena2012,Szymaniec2016,Levi2014}, reaching fractional frequency uncertainties at the level of $1$--$2 \times 10^{-16}$ limited by systematic shifts.

\item A $^{87}$Sr optical clock. Clouds of about $10^4$ atoms can be trapped in an optical lattice potential tuned close to a `magic wavelength' at 813~nm, where the differential polarisability between ground and excited states is zero.  The $^{87}$Sr clock has very small sensitivity factors, $K_{\alpha} = +0.06$ and $K_{\mu} = 0$, useful when comparing against clocks with larger sensitivities.  The current state-of-the-art for the $^{87}$Sr optical lattice clock~\cite{Bothwell2019} has an estimated fractional frequency uncertainty from systematic shifts of $2.0 \times 10^{-18}$.  

\item A  $^{171}$Yb$^+$ optical clock. This clock features an octupole transition that is very sensitive to variations of $\alpha$ with $K_{\alpha} = -5.95$. The current state-of-the-art for the E3 transition~\cite{Sanner2019} has an estimated fractional frequency uncertainty from systematic shifts of $2.7 \times 10^{-18}$.  

\item A CaF molecular lattice clock. This clock will be based on the fundamental vibrational transition in CaF, which has $K_\mu=0.5$. The main ideas for such a clock were presented by Kajita \cite{Kajita2018}. The ultracold molecules will be loaded into an optical dipole trap~\cite{Anderegg2018} or an optical lattice. Through a careful choice of transition, Zeeman, dc Stark and ac Stark shifts all cancel to high accuracy. In the optical lattice, the molecules are deep in the Lamb-Dicke regime which eliminates first-order Doppler shifts. A 3D lattice also eliminates collisional shifts. We estimate that systematic shifts can be controlled at the level of $\simeq8\times 10^{-18}$.

\item A molecular N$^+_2$ clock. The vibrational clock transition has a sensitivity of K$_\mu = 0.5$ and systematic shifts which are comparable with the current best optical clocks~\cite{Kajita} and facilitate frequency measurements at an uncertainty below $10^{-18}$. The molecular ions need to be trapped alongside atomic ions for sympathetic cooling and state detection. We evaluate that the fractional frequency uncertainty from systematic shifts for $N_2^+$ is $\simeq4\times 10^{-18}$ under conditions that can be easily reached in current experiments.

\item A Cf highly charged ion clock. Optical transition with values of $|K_\alpha|\simeq45$ are predicted to exist in the ionization states Cf$^{15+}$ and Cf$^{17+}$. Clocks based on highly-charged ions are expected to be less sensitive to external perturbations, and with the information at hand, in principle it would be possible to reach fractional frequency uncertainty on the order of $10^{-19}$ for both ionization states \cite{Cfclock}. Additionally, the possibility of realizing a dual clock co-trapping Cf$^{15+}$ and Cf$^{17+}$ is particularly appealing. 

\end{itemize}

\section{Summary}
\label{sec:conc} 

The QSNET project aims at realising a network of clocks with different sensitivities to variations of $\alpha$ and $\mu$. Such a network will explore large uncharted territories of the dark sector, and has the potential to impose new constraints over many models and theories, widening our understanding of the physics that governs the Universe. More specifically, QSNET will be sensitive to slow drifts of $\alpha$ and $\mu$, with relevance for, e.g., dark energy models and models that predict cosmological evolution of fundamental constants, fast oscillations of $\alpha$ and $\mu$, that can be linked to, e.g., virialised dark matter scalar fields, and transient events due to kinks or topological defects in dark matter fields.

\subsection*{Acknowledgments}
This work was supported by STFC and EPSRC under grants ST/T00598X/1, ST/T006048/1, ST/T006234/1, ST/T00603X/1,ST/T00102X/1,ST/S002227/1. We also acknowledge the support of the UK government department for Business, Energy and Industrial Strategy through the UK national quantum technologies programme.


\end{document}